\documentclass[prb,aps,twocolumn,floatfix,showpacs]{revtex4}
\usepackage{bm}
\usepackage{amssymb}
\usepackage{graphicx}

\begin{document}

\title{Hydrodynamic relation in 2D Heisenberg antiferromagnet in a field}

\author{A. L. Chernyshev}
\affiliation{Department of Physics, University of California, Irvine, 
California 92697, USA\\
 and Max-Planck-Institut f\"ur Physik komplexer Systeme,
01187 Dresden, Germany }

\author{M. E. Zhitomirsky} 
\affiliation{
Commissariat \`a l'Energie Atomique, DSM/INAC/SPSMS, F-38054 Grenoble,
France}

\date{\today}

\begin{abstract}
The spin-stiffness $\rho_s$ of a 2D Heisenberg 
antiferromagnet depends non-analytically on external magnetic field.
We demonstrate that the hydrodynamic relation between $\rho_s$, 
the uniform susceptibility $\chi$, and the spin-wave velocity $c$ 
is not violated by such a behavior because similar non-analytic terms 
from all three quantities mutually cancel out.
In this work, explicit expressions for the field-dependent spin stiffness 
and for the magnon velocity of the 2D square lattice antiferromagnet 
are obtained by direct calculation to order $1/S$ and in the whole 
range of magnetic fields. 
\end{abstract}

\maketitle

The effective description of spin-waves in the Heisenberg and 
easy-plane antiferromagnets by a hydrodynamic theory goes back to 
the  work by Halperin and Hohenberg. \cite{HH}
Such a description implies the following hydrodynamic relation
\begin{eqnarray}
\label{HR}
\frac{\chi\, c^2}{\rho_s}=1 ,
\end{eqnarray}
between the susceptibility $\chi$, the spin-wave
velocity $c$, and the spin stiffness $\rho_s$.
The importance of an independent verification of such a relation using 
direct microscopic calculations has been recognized and 
received a significant attention in the past.\cite{Hamer} 
Corresponding calculations confirming the validity of such a relation
for the 2D square-lattice Heisenberg antiferromagnet
(HAF) have been carried out in the early 1990s using the
spin-wave theory\cite{Igarashi,Hamer}  to orders $1/S^2$ and $1/S^3$.
Numerical studies\cite{Huse_Singh,Hamer,Einarsson} of the $S=1/2$ case 
of the same model have also given a strong support of the relation 
(\ref{HR}). While initial interest in this problem was motivated 
by the large-$J$ high-T$_c$ materials, more recently,  synthesis 
of small-$J$ quantum antiferromagnets\cite{Landee} 
has generated significant interest in the effects of external 
magnetic field in the properties of the HAFs, the regime that was 
previously unreachable.

Uniform magnetic field lowers the full rotational symmetry of 
the Heisenberg model to $O(2)$, making it equivalent to that of the 
the easy-plane antiferromagnets with the easy-plane of spin rotations 
perpendicular to the direction of the field. Note that the hydrodynamic
consideration of Ref.~\onlinecite{HH} is also valid for the
easy-plane antiferromagnets. Thus, at the first glance, it seems
natural to assume that the two hydrodynamic descriptions should
connect continuously. 
However, the situation is far less trivial 
as several quantities were shown to exhibit a non-analytic behavior in 
small fields. In the earlier work, Ref.~\onlinecite{Fisher_89},  
field dependence of the ground-state energy and 
susceptibility was discussed for the non-linear ${\bm \sigma}$-model. 
The non-analytic field-dependent corrections have been found
in the dimensions $D\leq 3$.  A subsequent independent study,
Ref.~\onlinecite{Zh_Nikuni}, obtained the same non-analytic 
behavior in small fields in the framework of the spin-wave theory. 
The recent work, Ref.~\onlinecite{Kopietz},
used a hybrid $1/S$-expansion$-{\bm \sigma}$-model approach to 
demonstrate that the spin-wave velocity in a 2D antiferromagnet also has a 
non-analytic dependence on the field, $c(H)-c(0)\propto |H|$ 
in the first $1/S$ order. 
Recent studies of the combined effects of the 
Dzyaloshinskii-Moriya and uniform magnetic field in the spectrum
and the ground-state properties\cite{Mila,Chernyshev_05} 
of the 2D HAFs have also found non-analytic dependencies that are 
related to the ones discussed here.

The origin of the non-analytic behavior can be traced to the 
field-induced gap in one of the Goldstone modes.
External field creates the so-called uniform-precession mode, 
which corresponds 
to the precession of the field-induced magnetization around the 
field direction with the energy equal to $H$. When the field is small, 
the mode is almost gapless and contributes to the fluctuation corrections
to various quantities. 
These fluctuations may, potentially, induce non-hydrodynamic corrections 
in the corresponding $1/S$ order of the theory. 
Thus, the validity of the 
relation (\ref{HR}) in a field has to be verified.

In the case of the square-lattice HAF,
out of three constants needed for the hydrodynamic relation it is
only the spin stiffness for which the presence or absence of the 
non-analytic terms in the field-dependence remains unknown.
In this work we carry out direct analytical calculations 
of $\rho_s$ to the necessary order in both $1/S$ and $H$ to: 
(i) identify such non-analytic terms, and 
(ii) verify that the non-analytic behavior of all three quantities 
{\it does not} lead to the violation of the hydrodynamic
relation (\ref{HR}). In the course of such derivation, 
we also obtain an analytic expression for the  
spin stiffness  to order $1/S$ and for all ranges of the field.
In addition, the non-analytic behavior of the spin-wave
velocity, previously obtained in Ref. \onlinecite{Kopietz} by a 
hybrid $1/S-{\bm\sigma}$-model approach, is confirmed within the
framework of the spin-wave theory and the compact analytic
expression for the velocity renormalization is obtained
for an arbitrary value of the field. 

We would like to make a separate note on the recent work,
Ref.~\onlinecite{Lauchli}, that combines a thorough numerical
investigation of the static and dynamic properties of the $S=1/2$ 
square-lattice HAF in a field with the spin-wave analysis of the
problem and provides a comprehensive comparison of the results. 
While in this work the spin stiffness is evaluated within 
the $1/S$ spin-wave approximation, it is done by numerical
differentiation of the energy with respect to the twist angle, and
no non-analytic behavior of $\rho_s$ vs.\ $H$ is discussed. Also,
the hydrodynamic relation is used to
provide a better estimate of the spin-wave velocity within the $1/S$
approach, but the validity of it is not verified.\cite{Lauchli} 
\begin{figure}[t]
\includegraphics[width=1\columnwidth]{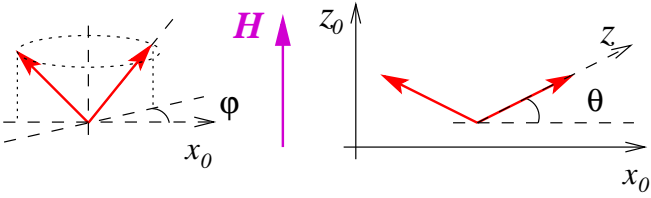}
\caption{(Color online) Left: 
angle of twist $\varphi$ in the $x_0-y_0$ plane of 
spins in one sublattice with respect to the other. Right:
field-induced canting by the angle $\theta$.}
\label{angles}
\end{figure}
 
We consider the spin-$S$ HAF on the square 
lattice in an external field along the $z_0$ axis of the laboratory 
reference frame with the Hamiltonian given by
\begin{eqnarray}
\hat{\cal H}  =  
J \sum_{\langle ij\rangle}  {\bf S}_i\cdot {\bf S}_j -
H\sum_{i}  S^{z_0}_i\ ,
\label{H}
\end{eqnarray}
where $\langle ij\rangle$ refer to the nearest-neighbor bonds.
To study the spin stiffness, the Hamiltonian should be modified 
to introduce a twist angle between spins, rigidity to which should 
yields the stiffness directly. One of the
prescriptions\cite{Huse_Singh} is to twist 
spins in every second row by the fixed angle $\varphi$. 
Another, intuitively more symmetric approach is to twist 
all the spins in one sublattice relative to the other.\cite{Lauchli}
In the latter method the twist energy is two times larger than
in the former case because every spin has twice as many nearest 
neighbors that are twisted. 
For the Heisenberg model on a bi-partite lattice in zero field the 
direction of such a uniform twist is arbitrary.
In the case of a non-zero external field, such a twist should 
be made in the plane perpendicular to the direction of the field, 
that is, in the $x_0-y_0$ plane, see Fig.~\ref{angles}.
Thus, using the sublattice twist with a small angle $\varphi\ll 1$, 
the modified Hamiltonian  reads:
\begin{eqnarray}
\hat{\cal H}  \approx  
J \sum_{\langle ij\rangle}  {\bf S}_i\cdot {\bf S}_j -
\frac{J\varphi^2}{2} \sum_{\langle ij\rangle}  
{\bf S}^\perp_i\cdot {\bf S}^\perp_j-
H\sum_{i}  S^{z_0}_i\ ,
\label{H1}
\end{eqnarray}
where ${\bf S}^\perp_i=(S_i^{x_0},S_i^{y_0},0)$ and we have omitted the
terms that are {\it linear} in $\varphi$, as they either vanish or 
contribute to the $\rho_s$-term only in the higher ($1/S^2$)
order.\cite{Hamer,Igarashi} As such, the 
Hamiltonian (\ref{H1}) contains all the necessary terms to study both 
the classical limit of the model and the $1/S$ fluctuation corrections to it. 

To study quantum fluctuations around the classical spin configuration
it is convenient to transform spins to ``rotating'' local reference
frames in which the quantization axis $z$ is along the classical
spin direction.\cite{Zh_Nikuni,field,Chernyshev_06}
Magnetic field cants spins toward its direction as is shown in 
Fig.~\ref{angles}. Assuming that the spins lie in
the $x$--$z$ plane we perform transformation from the laboratory frame 
$(x_0,z_0)$ into the rotating frame $(x,z)$:\cite{Zh_Nikuni,field}
\begin{eqnarray}
\label{transformation}
S_i^{z_0} & = & S_i^z \sin\theta - e^{i{\bf Q}r_i}S_i^x \cos\theta  
\ , \\
S_i^{x_0} & = & e^{i{\bf Q}r_i} S_i^z \cos\theta + S_i^x \sin\theta  \ ,
\ \ S_i^{y_0} =  S_i^y\ ,
\nonumber
\end{eqnarray}
where ${\bf Q}=(\pi,\pi)$ is the ordering wave-vector, and canting 
angle is as shown in  Fig.~\ref{angles}.  
The spin Hamiltonian (\ref{H1}) 
in the local coordinate system (\ref{transformation}) 
takes the  form:
\begin{eqnarray}
\hat{\cal H} & \approx & 
J \sum_{\langle ij\rangle} \Bigl[ S_i^yS_j^y  - 
\cos 2\theta \left(S^z_iS^z_j +  S^x_iS^x_j\right)     
\nonumber  \\
& & \mbox{} - \frac{\varphi^2}{2}
\left(S_i^yS_j^y +  S^x_iS^x_j\sin^2\theta -  S^z_iS^z_j\cos^2\theta \right)
\Bigr]\, ,
\label{H2}\\
& &-H\sin\theta\sum_{i}  S^{z}_i\ ,\nonumber
\end{eqnarray}
where, again, the terms that are not contributing to the harmonic
approximation are omitted. 

At the first glance, the spin stiffness can be defined from 
the averaging of the 
second line in Eq.~(\ref{H2}) over the ground state. However, the situation 
is slightly more complex as the field-induced canting angle $\theta$, which 
should be found from the minimization of the classical energy in 
(\ref{H2}),\cite{Zh_Nikuni} also depends on the twist angle.\cite{Lauchli} 
Performing such a minimization for (\ref{H2}),\cite{Zh_Nikuni} one obtains:
\begin{eqnarray}
\sin \theta = h\left(1+\frac{\varphi^2}{4}\right)   ,  
\label{theta}
\end{eqnarray}
where the terms of higher order in the twist angle are truncated and 
the dimensionless variable $h=H/(8JS)$, the field normalized to the
saturation field $H_s=8JS$ at which spins become fully aligned, 
is introduced. With the help of (\ref{theta}) one can eliminate $\theta$ in
(\ref{H2}) to obtain
\begin{eqnarray}
\hat{\cal H} & \approx & \hat{\cal H}_{\varphi=0}+
\hat{\cal H}_{\rho_s}\ ,
\label{H3}
\end{eqnarray}
where $\hat{\cal H}_{\varphi=0}$ contains no twist angle 
and $\hat{\cal H}_{\rho_s}$ is given by:
\begin{eqnarray}
\label{H4}
\hat{\cal H}_{\rho_s} & = & \frac{J\varphi^2}{2}
\bigg\{\sum_{\langle ij\rangle} \Bigl[ (1+h^2) S_i^zS_j^z+ 
h^2 S^x_iS^x_j -  S^y_iS^y_j\Bigr]\nonumber\\
&& {\phantom{\frac{J\varphi^2}{2}\sum_{\langle ij\rangle}}} 
-4Sh^2\sum_{i} S^{z}_i\bigg\}\ ,
\end{eqnarray}

The subsequent 
treatment of the  Hamiltonian 
$\hat{\cal H}_{\varphi=0}$ involves standard 
bosonization of spin operators via the Holstein-Primakoff 
transformation to the first $1/S$ order:
\begin{eqnarray}
S_i^z=S- a^\dagger_ia_i\ , \ \ 
S_i^-\approx a^\dag\sqrt{2S}\ , \ \ S_i^+=(S_i^-)^\dag\ ,
\label{HP}
\end{eqnarray} 
which is followed by the  Bogolyubov transformation.\cite{Zh_Nikuni}
This yields the linear spin-wave theory Hamiltonian:
\begin{eqnarray}
\label{H5}
\hat{\cal H}_{\varphi=0}  \approx  E_{\rm GS}+4JS\sum_{\bf k} \omega_{\bf k} 
\alpha^\dag_{\bf k} \alpha_{\bf k}\ , 
\end{eqnarray}
see Ref.~\onlinecite{Zh_Nikuni} for details. The 
dimensionless frequency is:
\begin{equation}
\omega_{\bf k} = 
\sqrt{(1+\gamma_{\bf k})(1-(1-2h^2)\gamma_{\bf k})} \ ,
\label{nu}
\end{equation} 
and $\gamma_{\bf k}=(\cos k_x+ \cos k_y)/2$.
\begin{figure}[t]
\includegraphics[width=1\columnwidth]{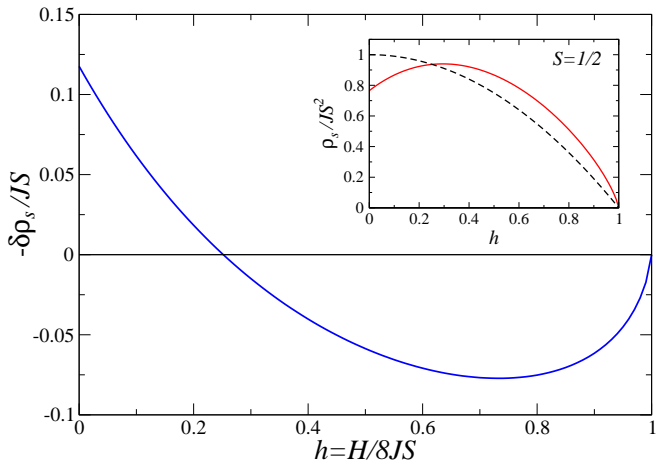}
\caption{(Color online) $1/S$ quantum correction to the spin stiffness
  $-\delta\rho_s/JS$ as a function of the field $h$.  
Inset: $\rho_s/JS^2$ vs $h$ for $S=1/2$, dashed line is 
$\rho_s^{\rm cl}/JS^2$ vs $h$.}
\label{drho}
\end{figure}

After diagonalization of $\hat{\cal H}_{\varphi=0}$, 
the spin stiffness can be found as a coefficient in front of 
$\varphi^2$ in the twist part $\hat{\cal H}_{\rho_s}$ 
of the Hamiltonian (\ref{H4}) by averaging the spin operators 
over the spin-wave ground state and keeping terms to $1/S$
order. Having in mind the extra factor of 2 in the sublattice twist
approach, this finally yields:
\begin{eqnarray}
\label{H6}
\delta E_\varphi=\left\langle\hat{\cal H}_{\rho_s}\right\rangle 
=\varphi^2 N \rho_s\ ,
\end{eqnarray}
where $\rho_s=\rho_s^{\rm cl} +\delta\rho_s$ with the classical and 
quantum contributions given by
\begin{eqnarray}
\label{rhos1}
\rho_s^{\rm cl}&=&JS^2\left(1-h^2\right)\ , \\
\delta\rho_s&=&-JS\left[2n-\Delta(1+h^2)+m(1-h^2)\right]\ ,
\nonumber
\end{eqnarray}
where we use the following Hartree-Fock averages of the 
two-boson operator combinations:
\begin{eqnarray}
\label{HF}
n &=& \langle a^\dagger_i a_i\rangle 
= \frac12\sum_{\bf q} 
\Bigl[\frac{1+h^2\gamma_{\bf q}}{\omega_{\bf q}}-1\Bigr]\ 
, \nonumber\\
\Delta &=& \langle a_i a_j\rangle= 
\frac12(1-h^2)\sum_{\bf q}\frac{\gamma_{\bf q}^2}{\omega_{\bf q}} \ , \\ 
 m &=& \langle a^\dagger_i a_j\rangle=
\frac12\sum_{\bf q} 
\frac{\gamma_{\bf q}+h^2\gamma_{\bf q}^2}{\omega_{\bf q}} \ , 
\nonumber
\end{eqnarray}
The above result (\ref{rhos1})
gives $\rho_s$ to the order $1/S$ and for the fields anywhere between 
zero and the saturation value. 
At $H=0$ the expression for $\rho_s$ in (\ref{rhos1}) and 
(\ref{HF}) coincides with the known zero-field formula for the
Heisenberg model. \cite{Igarashi,Hamer} 
The field-dependence of the 
quantum correction to the spin stiffness $\delta\rho_s$ is shown
in Fig.~\ref{drho}.  
The inset presents $\rho_s$ for
the spin-1/2 case. One of the interesting
observations is that $\delta\rho_s$ changes sign as a function of 
the field. It also exhibits a singular behavior in the derivative 
as $h\rightarrow 1$, similar to the one discussed before for the
magnetization,\cite{Zh_Nikuni} and is related to the logarithmically
vanishing scattering amplitude in the dilute 2D gas of bosons. The linear
(non-analytic) field-dependence at small field is also clear from
Fig.~\ref{drho}. 

With the expressions (\ref{rhos1}) and (\ref{HF}) at hand, one can now 
study the field dependence of $\rho_s$ at $h\rightarrow 0$. After some
algebra one finds:
\begin{eqnarray}
\label{rhos2}
\rho_s=\rho_s^{H=0}-JS\sum_{\bf k}\left(\frac{1}{\omega_{\bf k}}-
\frac{1}{\omega^{H=0}_{\bf k}} \right) + {\cal O}(h^2).
\end{eqnarray}
It is easy to see that due to the field-induced 
gap $\propto H$ in the magnon
spectrum the fluctuation terms like the one in (\ref{rhos2}) 
are yielding corrections $\propto |H|$ in 2D. 
Some further algebra gives:
\begin{eqnarray}
\rho_s&=&JS^2\left(Z_{\rho_s}+\frac{2}{\pi S}\,|h|\right)+ 
{\cal O}(h^2, 1/S^2)\nonumber\\
&\approx& \rho_s^{H=0}\left(1+\frac{2}{\pi S}\,|h|\right), 
\label{rhos3}
\end{eqnarray}
where $\rho_s^{H=0}$ contains zero-field, $1/S$ renormalization factor
$Z_{\rho_s}$, Ref.~\onlinecite{Hamer}, 
and the last expression is obtained within the 
same $1/S$ accuracy.

\begin{figure}[t]
\includegraphics[width=1\columnwidth]{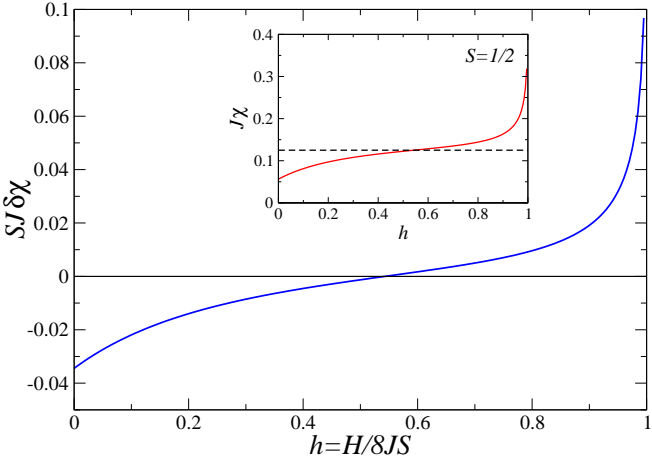}
\caption{(Color online) $1/S$ quantum correction to the susceptibility
  $JS\delta\chi$ as a function of the field $h$.  
Inset: $J\chi$ vs $h$ for $S=1/2$, dashed line is 
$J\chi_{\rm cl}$.}
\label{dchi}
\end{figure}

For completeness, we also list  the corresponding $1/S$
expressions for the susceptibility.
Magnetization of the square-lattice HAF is $M=M_{\rm cl}+\delta M$
with the classical part and the quantum correction given by:\cite{Zh_Nikuni}
\begin{eqnarray}
M_{\rm cl}=Sh, \ \ \ \ \delta M=-h(m+\Delta).
\label{dM}
\end{eqnarray}
Using $\chi=\partial M/\partial H$ yields:
\begin{eqnarray}
\chi=\chi_{\rm cl}+\delta\chi=\frac{1}{8J}
\Bigl[1-\frac{1}{S}\bigl(m+\Delta - h^2 I_1\bigr)\Bigr],
\label{d_chi}
\end{eqnarray}
where $I_1$ stands for the integral
\begin{equation}
\label{Ichi}
I_1 = \sum_{\bf q} \frac{\gamma_{\bf q}^2(1+\gamma_{\bf q})^2}
{\omega_{\bf q}^3} \ .
\end{equation}
At small fields, the same algebra as above gives:
\begin{eqnarray}
\chi&=&\frac{1}{8J}\Bigl(Z_{\chi}+\frac{4}{\pi S}\,|h|\Bigr)+ 
{\cal O}(h^2, 1/S^2)\nonumber\\
&\approx& \chi^{H=0}\Bigl(1+\frac{4}{\pi S}\,|h|\Bigr).
\label{chi}
\end{eqnarray}
Figure \ref{dchi} shows the field-dependence of the 
quantum correction to the susceptibility $\delta\chi$. 
The inset presents $\chi$ for
the case of $S=1/2$. Similarly to $\delta\rho_s$, $\delta\chi$ 
changes sign as a function of 
the field.  As is discussed in Ref.~\onlinecite{Zh_Nikuni}, $\chi$
has a singular logarithmic behavior at $h\rightarrow 1$ and the linear,
non-analytic field-dependence at $h\rightarrow 0$ is also clearly seen
in Fig.~\ref{dchi}. 

For the square lattice HAF in external
magnetic field the energy of magnons to 
the first-order of the $1/S$ expansion is given by\cite{field}
\begin{equation}
\bar{\varepsilon}_{\bf k} = 4JS\omega_{\bf k}
+ \delta\varepsilon^{(1)}_{\bf k} + \delta\varepsilon^{(2)}_{\bf k} \ ,
\end{equation}
where $\omega_{\bf k}$ is defined in Eq.~(\ref{nu}),  correction 
$\delta\varepsilon^{(1)}_{\bf k}$ 
includes the Hartree-Fock and the canting angle 
renormalizations 
\begin{eqnarray}
\label{e1}
\delta\varepsilon^{(1)}_{\bf k} & =&  \frac{4J}{\omega_{\bf k}}
\Bigl\{ \Delta\! -\! n \!+ \!h^2 (\Delta \!+ \!m)\left[1 
-2\gamma_{\bf k} (1\!-\!h^2)\right] \\
& + & \gamma^2_{\bf k}
\bigl[n - \Delta(1\!-\!h^2\!+\!2h^4) - mh^2(3\!-\!2h^2)
\bigr]\Bigr\} \nonumber
\end{eqnarray}
and $\delta\varepsilon^{(2)}_{\bf k}$ is the one-loop contributions
from the three-magnon coupling:  
\begin{eqnarray}
\label{e2}
\delta\varepsilon^{(2)}_{\bf k}  &=&  -4Jh^2(1-h^2)   \\
&\times&
\sum_{\bf q}
\biggl[ \frac{\widetilde{\Gamma}_1({\bf k},{\bf q})^2}
{\omega_{\bf q}\!+\omega_{\bf k-q+Q}\!-\omega_{\bf k}}
+ \frac{\widetilde{\Gamma}_2({\bf k},{\bf q})^2}
{\omega_{\bf k}\!+\omega_{\bf q}\!+\omega_{\bf k+q-Q}}\biggr]\nonumber
\end{eqnarray}
Explicit expressions for $\widetilde{\Gamma}_1({\bf k},{\bf q})$ and
$\widetilde{\Gamma}_2({\bf k},{\bf q})$ are given in 
Ref.~\onlinecite{field}. After some algebra, the $1/S$ correction 
to the spin-wave velocity can be written as:
\begin{eqnarray}
\label{velocity2}
\frac{c-c_0}{c_0} &=& \frac{\Delta(1-h^2\!+h^4) - n + 
m h^2(2-h^2)} {S(1-h^2)}  \\
& & \mbox{} -\frac{h^2}{2S}\,I_1 -\frac{2h^2}{S}\!\left.\biggl(\frac{I_2({\bf k}) - 
4\Delta}{k^2}\biggr)\right|_{k\rightarrow 0} \ , \nonumber
\end{eqnarray}
where the bare spin-wave velocity is  
$c_0^2=8J^2S^2(1-h^2)$ 
and the ${\bf k}$-dependent function in the last term is: 
\begin{equation}
\label{I4}
I_2({\bf k}) = 2(1-h^2) \sum_{\bf q} \gamma_1\,\frac{
\gamma_1 \alpha_1 \beta_2
+ \gamma_2 \omega_1 \omega_2}
{\omega_1\omega_2 (\omega_1+\omega_2)} \ ,
\end{equation}
with $\alpha_1=(1+\gamma_1)$, 
$\beta_2=\left[1-(1-2h^2) \gamma_2\right]$, 
and  $1,2 = {\bf q},{\bf q-k}$. 

\begin{figure}[t]
\includegraphics[width=1\columnwidth]{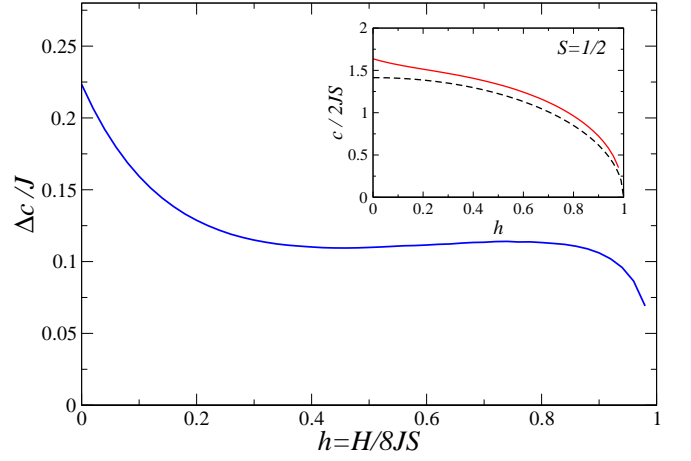}
\caption{(Color online) $1/S$ quantum correction to the spin-wave velocity
  $\Delta c/J$ as a function of the field $h$.  
Inset: $c/2JS$ vs $h$ for $S=1/2$, dashed line is $c_0/2JS$ vs $h$.}
\label{dc}
\end{figure}

The field-dependence of the quantum correction to the spin-wave 
velocity $\Delta c$, obtained by numerical evaluation of the 
integrals in Eq.~(\ref{velocity2}), is shown in Fig.~\ref{dc}. 
The inset presents the normalized magnon velocity for
$S=1/2$. The linear
(non-analytic) field-dependence at small field is clearly visible 
in Fig.~\ref{dc}. Behavior of $\Delta c$ at $h\rightarrow 1$ is also
singular, similarly to other quantities.
It is interesting to note that the correction to the spin-wave
velocity $\Delta c$ is almost flat for $0.2\alt h\alt 0.9$. 

It is easy to see that the first Hartree-Fock term in
Eq.~(\ref{velocity2}) does not contribute to the anomalous non-analytic
field dependence. After some more algebra, one can show that the same
is true  for the last term in (\ref{velocity2}). 
The second term in Eq.~(\ref{velocity2}), on the other hand, 
yields at $h\rightarrow 0$:
\begin{equation}
\frac{\Delta c}{c_0} \approx - \frac{2h^2}{S} \sum_{\bf q} 
\left(\frac{1}{(\omega_{\bf q})^3} -\frac{1}{(\omega^{H=0}_{\bf q})^3} \right)
 = - \frac{|h|}{\pi S} \ .
\end{equation}
This gives the same result as in Ref.~\onlinecite{Kopietz}:
\begin{equation}
\label{c}
c^2\approx c^2_{H=0}\left(1-\frac{2}{\pi S}\,|h|\right).
\end{equation}

\begin{figure}[t]
\includegraphics[width=1\columnwidth]{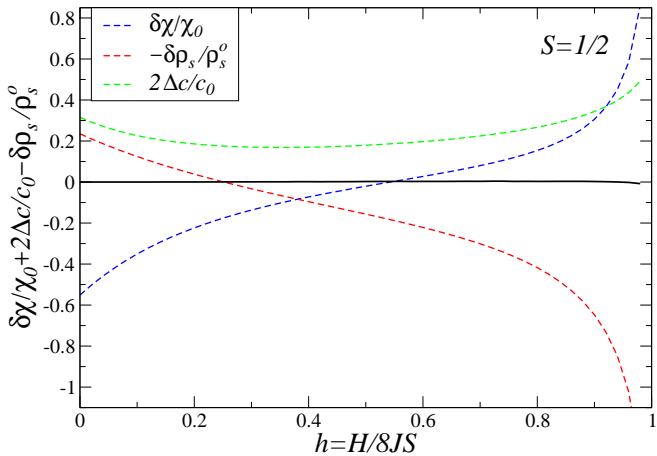}
\caption{(Color online) Cancellation of the
$1/S$ quantum corrections in the hydrodynamic relation (\ref{HR1})
as a function of the field $h$ (solid line). Dashed lines
show contributions of individual terms in (\ref{HR1}).
}
\label{hydro}
\end{figure}

Combining the expressions for the small-field expansion 
of all three quantities, $\rho_s$, $\chi$, and $c$ from 
Eqs.~(\ref{rhos3}), (\ref{chi}), and
(\ref{c}), one can easily see that the hydrodynamic relation 
(\ref{HR}) is obeyed as all the non-analytic terms explicitly 
cancel each other in the leading
order in $h$. Moreover, such a verification of the relation 
(\ref{HR}) can be extended to an arbitrary field. 
Expanding the hydrodynamic relation (\ref{HR}) to $1/S$ order and
observing that it is fulfilled at the classical
level, $\chi_{\rm cl}\, c_0^2/\rho_s^{\rm cl}=1$, 
one concludes that for the
hydrodynamic relation to exist 
the quantum corrections from all three quantities 
must cancel each other at any field in each order of $1/S$. For the
$1/S$ corrections this leads to
\begin{eqnarray}
\label{HR1}
\frac{\chi\, c^2}{\rho_s}-\frac{\chi_{\rm cl}\, c_0^2}{\rho_s^{\rm cl}}
\approx\left(\frac{\delta\chi}{\chi_{\rm cl}}+
2\frac{\Delta c}{c_0}-\frac{\delta\rho_s}{\rho_s^{\rm cl}}\right)=0 \ .
\end{eqnarray}
Numerical verification of the above relation is made using
expressions (\ref{rhos1}), 
(\ref{d_chi}), and (\ref{velocity2}) and is presented by
the solid line in Fig.~\ref{hydro}. The dashed lines
show contributions of individual terms in (\ref{HR1}). One can conclude
that the cancellation takes place for all values of $0<h<1$.

Having in mind the relation (\ref{HR1}), we can now
obtain a much simpler expression for the spin-wave velocity renormalization
in magnetic field:
\begin{eqnarray}
\label{velocity3}
\frac{c-c_0}{c_0} =\frac{1}{S} \Bigl(\frac{\Delta - n} 
{1-h^2} - \frac{h^2}{2} I_1 \Bigr)\ ,
\end{eqnarray}
where $I_1$ is defined in Eq.~(\ref{Ichi}).

Altogether, we have confirmed the validity of the hydrodynamic
relation for the 2D Heisenberg antiferromagnet in a uniform
field. Despite the appearance of the non-analytic terms in the 
field-dependence of all key quantities due to quantum 
fluctuation involving
small field-induced gap, they are not sufficient to violate
such a relation. We have obtained  expressions for the spin-stiffness 
$\rho_s$ and for the spin-wave velocity $c$ 
for the square-lattice HAF, valid to the first-order in $1/S$ and for
the whole range of magnetic fields. The non-analytic 
field-dependence of $c$, 
previously obtained by a hybrid $1/S$-expansion$-{\bm \sigma}$-model
approach, is verified using the more conventional spin-wave theory.


We are grateful to D. V. Efremov for useful discussion and to  
P. Kopietz for communications and discussions. 
Part of this work has been done at the Max-Plank Institute for Complex
Systems which we would like to thank for hospitality.
This work was supported by DOE under grant DE-FG02-04ER46174 (A.L.C.)
and by the visiting professorship at the Institute for Solid State Physics,
University of Tokyo (M.E.Z.).



\end{document}